\documentstyle[11pt,newpasp,twoside]{article}

\def\edcomment#1{\iffalse\marginpar{\raggedright\sl#1\/}\else\relax\fi}

\reversemarginpar

\begin{document}

\title{Space Studies of the Black-Drop Effect at a Mercury Transit}

 \author{Glenn Schneider}

\affil{Steward Observatory, 933 N. Cherry Ave., University of Arizona, Tucson,  Arizona  85721}

\author{Jay M. Pasachoff}

\affil{Williams College-Hopkins Observatory, 33 Lab Campus Dr., Williamstown,  MA  01267}

\author{Leon Golub}

\affil{Smithsonian Astrophysical Observatory, Mail Stop 58, 60 Garden St., Cambridge, MA 02138}

\begin{abstract}
Transits of Mercury and Venus across the face of the Sun are rare.
The 20th century had 15 transits of Mercury and the 21st century will
have 14, the two most recent occuring on 15 November 1999 and 7 May 2003.
We report on our observations and analyses of a black-drop effect at the
1999 and 2003 transits of Mercury seen in high spatial resolution
optical imaging with NASA's Transition Region and Coronal Explorer (TRACE)
spacecraft. We have separated the primary contributors to this effect,
solar limb darkening and broadening due to the instrumental point spread
function,  for the 1999 event. The observations are important for understanding
historical observations of transits of Venus, which in the 18th and 19th centuries
were basic for the determination of the scale of the solar system.  Our
observations are in preparation for the 8 June 2004 transit of
Venus, the first to occur since 1882. Only five
transits of Venus have ever been seen -- in 1639, 1761, 1769, 1874, and
1882.   These events occur in pairs, whose members are separated by 8
years, with an interval between pairs of 105 or 122 years.  Nobody
alive has ever seen a transit of Venus.
\end{abstract}

\section{Motivation}
Historically, transits of Venus were the major method for hundreds of
years of determining the Astronomical Unit and thus the scale of the
solar system, given that Kepler's laws of 1609/1618 are mere
proportions.  Edmond Halley presented a method of determining the
A.U. by observing the durations of the chords across the Sun from a
number of different locations on Earth.  Accordingly, dozens of
expeditions from many countries travelled around the world for the
18th and 19th-century transits, most famously including the voyage of
Captain James Cook, who was sent to Tahiti to observe the 1769 event.

The accuracy of the measurements was severely impaired, however, by
the ``black-drop effect,'' in which the silhouette of Venus did not
separate cleanly from the limb of the Sun during its inner contact.
The timing accuracy was thus closer to a minute than to the expected
second or two.  As Schaefer (2001) has shown, many people have
mistakenly attributed, and continue to mistakenly attribute, the
black-drop effect to the atmosphere of Venus.

Our space observations of the transit of Mercury  have shown the
presence of a black-drop effect (Figure 1).  Since Mercury has no substantial
atmosphere and since the observations were taken from outside the
Earth's atmosphere, clearly the effect -- at least for Mercury -- has
causes other than due to a planetary atmosphere.  By implication,
Venus's black-drop effect would arise, at least in part, from similar
causes.

\section{Observations}
The Transition Region and Explorer spacecraft can observe the Sun in a
variety of ultraviolet wavelengths.  For our studies of the 1999
transit of Mercury, we used only data from its broad band white-light channel.
TRACE's 0.5 arc sec pixels, and temporally stable point-spread
function unaffected by the Earth's atmosphere,  give it spatial high resolution.
A black-drop effect was observed in all frames near the point of internal
tangency of the Mercurian and solar disks.

We discuss our calibrations, methods of analysis and extensive
modeling in Schneider, Pasachoff \& Golub (2001, 2004).  We found 
major contributors giving rise to the black-drop effect come
from two causes, that of the point-spread function of the telescope
and that of the solar limb darkening.  Removing those two
contributions left us with images of the limb of Mercury 
unaffected by the solar limb.

The data for the 1999 event were sent to Earth in lossless uncompressed
image format.  Though we had worked with the TRACE planning 
team for the 2003 event to obtain a white-light data set at higher
temporal cadance, those observations were downlinked in a compressed
format normally used for solar observations.  A irrecoverable loss
of image fidelity at the bottom end of the dynamic sampling range
resulted, and we were unable to recover sufficient accuracy to repeat our
earlier analysis.  We will make sure that the data for
the upcoming Venus transit are returned to Earth in uncompressed format. 

\section{Future Work}
Our immediate intention is to observe the 8 June 2004 transit of Venus
from TRACE in orbit and from telescopes on the ground.  Massive
world-wide efforts will take place to observe this Venus transit.  The
European Southern Observatory, for example, is coordinating a public
observation campaign.  The International Astronomical Union's
Commission on Education and Development has a Web site at
http://www.transitofvenus.info that lists past and future
observations, shows images, and provides links.
        We also intend to observe the transit of Mercury of 8 November
2006.  After that, the following transits of Mercury aren't until 9 May 2016 and 11 November 2019.

\begin{figure}[ht]
\vspace*{8.1 truecm}
\includegraphics{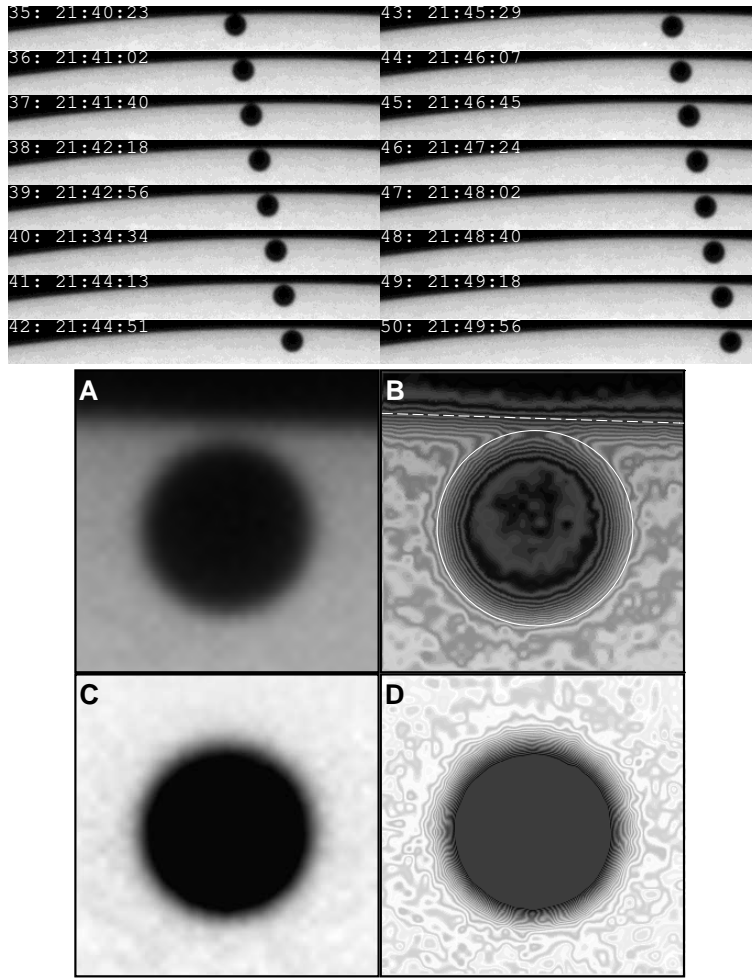}
\vspace{1.9 truein}
\caption[]{TRACE images. A-D 
from frame 50. A (image) \& B (5\% intensity contours; dashed line = solar limb, white circle = Mercury disk). Post-processed: C \& D (1/2\% contours) show no black-drop.}

\end{figure} 
\acknowledgments
This work was partially supported by NSF grant ATM-000545; Pasachoff's
work on eclipses has also been supported in part by NGS/CRE.
TRACE is supported by a NASA/GSFC contract to the Lockheed Martin
Corp. NICMOS IDT S/W (NASA grant NAG 5-3042) was used extensively.

\end{document}